\documentclass[conference]{IEEEtran}
\usepackage[noadjust]{cite}
\usepackage{amsmath,amssymb,amsfonts}
\usepackage{graphicx}
\usepackage{rotating}

\title{ Downscaling Extreme Rainfall Using Physical-Statistical Generative Adversarial Learning}

\author{\IEEEauthorblockN{Anamitra Saha}
\IEEEauthorblockA{\textit{Massachusetts Institute of Technology}\\
Cambridge, MA, USA \\
anamitra@mit.edu}
\and
\IEEEauthorblockN{Sai Ravela}
\IEEEauthorblockA{\textit{Massachusetts Institute of Technology}\\
Cambridge, MA, USA \\
ravela@mit.edu}
}

\begin{document}
\maketitle

\begin{abstract}
Modeling the risk of extreme weather events in a changing climate is essential for developing effective adaptation and mitigation strategies. Although the available low-resolution climate models capture different scenarios, accurate risk assessment for mitigation and adaption often demands detail that they typically cannot resolve. Here, we develop a dynamic data-driven downscaling (super-resolution) method that incorporates physics and statistics in a generative framework to learn the fine-scale spatial details of rainfall. Our method transforms coarse-resolution ($0.25^{\circ} \times 0.25^{\circ}$) climate model outputs into high-resolution ($0.01^{\circ} \times 0.01^{\circ}$) rainfall fields while efficaciously quantifying uncertainty. Results indicate that the downscaled rainfall fields closely match observed spatial fields and their risk distributions. 
\end{abstract}

\section{Introduction}
The susceptibility to extreme weather events such as cyclones will likely worsen in a changing climate under continued warming from greenhouse gas (GHG)  emissions~\cite{IPCC2021}. The adverse effects of weather extremes are broad, including but not limited to food security, urban infrastructure, public health, and ecological sustainability. Rising cyclones and extreme rainfall events, for example,  lead to more frequent and severe floods, causing economic damage and loss of human lives and livelihood~\cite{neumann2015joint, neumann2015risks}. Quantifying the risk of weather extremes in a changing climate is vital for optimizing climate change adaptation and mitigation strategies.

Conceptually, weather extremes are rare events within a non-stationary nonlinear stochastic process. Unfortunately, being few and far between, observational records of extreme events are often too short to determine future risk. Thus, the historical record alone is insufficient for modeling future risk. One could use physically-based numerical climate models to infer the frequency and severity of rare extremes in nature~\cite{john2022quantifying}. However, coarse-resolution models often fail to capture essential fine-scale detail, and they become computationally expensive at the needed high resolutions. Thus, there is strong interest in  {\it downscaling} (super-resolution) methods~\cite{wilbyDownscalingGeneralCirculation1997} with improved efficacy.

We posit that Machine Learning can provide the needed efficacy. The methodology is to learn a downscaling function using long-range simulations of a relatively small ensemble of high- and low-resolution climate model pairs under different climate scenarios. After that, downscaling many low-resolution climate model outputs to high-resolution fields is rapid. When trained with uncertainty-aware methods~\cite{trautner2020informative}, the downscaling function would also rapidly quantify uncertainty and risk. 

However, the learning approach also requires caution. Downscaling is limited to the finer model resolution, which is reasonable. Downscaling functions trained on one storyline and time window must generalize to other time windows and storylines for efficacy, which advances in transfer learning could address. Most pressingly, the high-resolution reference models are not ground truth. Resolving the compounding effect of model bias with downscaling imperfections is essential to improve trust and support the Machine Learning pathway. 

This paper takes an initial step in addressing these issues. In particular, we adopt a hindcasting approach using extreme events from years past in coarse climate model/reanalysis outputs and available fine-scale observations to train a downscaling function. Comparing downscaled climate model outputs from the more recent years with observations then tests performance. The approach is verifiable, and using actual data removes high-resolution model bias. Although the data processing may contain some bias, it is often a much better representation of the truth, and training also corrects low-resolution model bias. It is also immediately applicable. Training to the present time using a coarse climate model and projecting risk into the immediate future (inter-annual to decadal timescale) is key to many applications, such as underwriting insurance and disaster response planning. In a companion paper, Salas et al.~\cite{salas22} show high-resolution rainfall-coupled flood-loss modeling for interannual risk improves performance in shifting and expanding time window regimes.

Unfortunately, model and data-trained downscaling approaches have additional issues to resolve. Downscaling is already an ill-posed inverse problem, and the inability of ML solutions to satisfy basic physical principles without additional support complicates the situation. There is severe data paucity for rare extremes, and developing synthetic data augmentation strategies for “detail-preserving regularization” is challenging. An added challenge encountered with a “coarse-model input observed-data output” rainfall downscaling approach is that one-to-one correspondence between inputs and output training data doesn't typically exist. Additionally, missing/spurious events, timing errors, and intensity differences are possible. 

Here, we propose a novel dynamic data-driven downscaling approach~\cite{dddasbook} that overcomes these issues. While the methodology is extensible to many meteorological fields, the focus is rainfall extremes in the mid-latitudes, an important cause of inland flooding. 

\begin{figure*}[htbp]
\centering
\includegraphics[trim=0 0 0 78mm, scale=0.13]{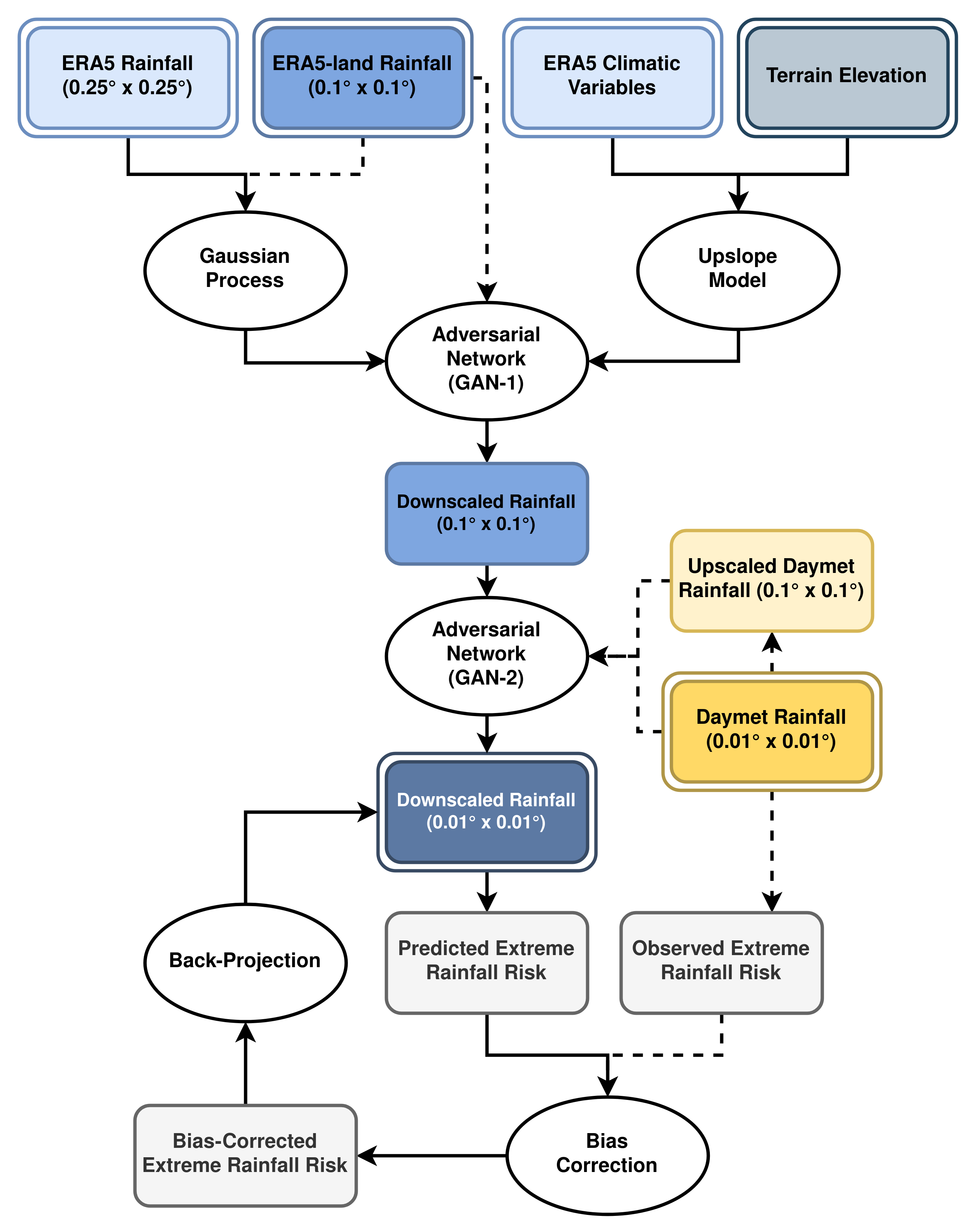}
\caption[Overview]{Overview of the proposed downscaling methodology. Coarse-resolution ($0.25^{\circ} \times 0.25^{\circ}$) climate model outputs are downscaled to fine-resolution ($0.01^{\circ} \times 0.01^{\circ}$) rainfall in two steps (GAN-1 and GAN-2). ERA5 rainfall is initially downscaled in the first step using conditional Gaussian Processes and combined with orographic rainfall estimates from the upslope model. The outputs pass to a Generative Adversarial Network (GAN-1), which produces fine-resolution rainfall fields. In the next step, another adversarial network (GAN-2) is trained on upscaled Daymet rainfall fields and then applied to the output of GAN-1 to produce even finer-resolution rainfall. Over a Validation period, rainfall return-period distributions computed from downscaled and observed rainfall fields train bias correction functions. Finally, Bias-corrected rainfall risk curves are back-projected onto rainfall fields. The white ellipses denote methods, and the colored double-boxes denote input or output to the models. The dotted arrows represent actions performed only during training and not in operation.}
\label{fig:overview}
\end{figure*}

Our approach consists of the following steps (see Figure~\ref{fig:overview}): 
\begin{enumerate}
    \item Physics: a simple upslope lift model captures the basic structure of rainfall fields and provides high-resolution estimates of orographic precipitation. A coarse climate model field input yields a relatively detailed orographic rainfall component.
    \item Dynamic Data-Driven Statistics: the training data (coarse model outputs and high-resolution observations) are indexed on a manifold and employ a conditional ensemble-based Gaussian process~\cite{trautner2020informative, ravela2016dynamically} to produce “first-guess” rainfall estimates.
    \item The “first-guess” and orographic rainfall estimates are input to the adversarial learning framework in the third step. We produce high-resolution deterministic rainfall from coarse model inputs with little training data by priming a two-stage Generative Adversarial Network (GAN) with physics and statistics.
    \item In the fourth step, we model and inject stochastic perturbations of residual excess rainfall into the deterministic output to produce an ensemble-based optimal estimate for bias correction~\cite{ravela2007fast, ravela2010realtime,trautner2020informative}. We show that the risk of extreme annual rainfall captured by downscaled predictions closely matches the observation.
\end{enumerate} 

Applying this method indicates that the four steps yield an algorithm that richly captures high-resolution spatial fields and matches the observed risk distributions. Using a climate/reanalysis model as input (ERA5) and observations (Daymet) as training output currently enables the inter-annual risk applications, with future applicability to climate risk using high-resolution model fields as training output. 

The rest of the paper is as follows. Section~\ref{section:relatedwork} discusses prior research on rainfall downscaling. Section~\ref{section:methods} explains the methodology. Section~\ref{section:results} presents results. Section~\ref{section:discussions} discusses the significance of this work and concludes the paper.

\section{Related Work}
\label{section:relatedwork}
Two major approaches for rainfall downscaling are found in the scientific literature: dynamical downscaling and statistical downscaling~\cite{wilbyDownscalingGeneralCirculation1997}. Dynamical downscaling involves embedding a relatively high-resolution regional climate model inside the coarse-resolution model~\cite{giorgi2009addressing}. The computationally expensive nature of the dynamical downscaling methods has led to the popularity of non-expensive statistical downscaling approaches, which entail establishing linear or nonlinear transfer functions from coarse-resolution predictor variables to fine-resolution rainfall. A plethora of statistical techniques for precipitation downscaling are available with varying degrees of success, such as Bias Correction and Spatial Disaggregation (BCSD)~\cite{wood2002long}, parametric regression~\cite{najafiStatisticalDownscalingPrecipitation2011}, kernel regression~\cite{kannanNonparametricKernelRegression2013}, non-homogeneous Markov model~\cite{mehrotra2005nonparametric}, Bayesian model averaging~\cite{zhang2015new} etc. Besides these statistical methods, neural networks such as multilayer perceptron~\cite{xuDownscalingProjectionMultiCMIP52020}, artificial neural network~\cite{schoof2001downscaling}, and quantile regression neural network~\cite{cannonQuantileRegressionNeural2011} are available for rainfall downscaling. Alternative machine learning approaches such as random forests~\cite{heSpatialDownscalingPrecipitation2016}, support vector machine (SVM)~\cite{tripathiDownscalingPrecipitationClimate2006}, dictionary learning~\cite{xu2020precipatch} and genetic programming~\cite{sachindraMachineLearningDownscaling2019} are also applicable to solve the downscaling problem.

With the advent of deep learning techniques, a new suite of ML-based approaches, such as Recurrent Neural Networks (RNN)~\cite{wangSequencebasedStatisticalDownscaling2020}, Long Short-Term Memory (LSTM)~\cite{miaoImprovingMonsoonPrecipitation2019}, autoencoder~\cite{vandalIntercomparisonMachineLearning2019}, U-net~\cite{shaDeepLearningBasedGriddedDownscaling2020} etc. have been available for rainfall downscaling. Because of their deep layered structure, deep learning methods are well-suited for extracting high-level feature representations from high-dimensional climate datasets. Several Convolutional Neural Network (CNN)-based approaches, developed for single image super-resolution, are also brought into the climate science domain, as they can explicitly capture the spatial structure of climate variables. Although the methods primarily apply to computer vision, insights also apply to the precipitation downscaling problem. Super-resolution CNN (SRCNN)~\cite{dong2015image} was one of the first successful approaches developed in this field. Many subsequent models, such as Very Deep Super-resolution (VDSR)~\cite{kim2016accurate}, Enhanced Deep Super-resolution (EDSR)~\cite{lim2017enhanced}, Deep Back-projection Network (DBPN)~\cite{haris2018deep} etc., were built upon it. Adversarial super-resolution method Super-resolution Generative Adversarial Network (SRGAN)~\cite{ledig2017photo} showed that GAN could better model the high-frequency distribution of the image and improve the sharpness and perceptual quality. Enhanced Super-resolution GAN (ESRGAN)~\cite{wang2018esrgan} improves upon this approach by providing a better GAN-loss formulation. These CNN and GAN-based methods have since emerged for various rainfall downscaling studies~\cite{vandalDeepSDGeneratingHigh2017, singh2019downscaling, watson2020investigating}.

Another direction involves physics-based approaches for rainfall downscaling, where an analytically estimated high-resolution orographic precipitation component augments climate model-simulated large-scale rainfall fields. The literature discusses the impact of orography on regional precipitation patterns and orographic precipitation modeling~\cite{roe2005orographic}. A simple model for orographic rainfall estimation would be the Upslope model~\cite{collier1975representation}, where we assume the condensation rate is proportional to the vertical wind velocity, and the condensed rainwater falls immediately to the ground. Considerations of downstream hydrometeor drift~\cite{sinclair1994diagnostic} may reduce these biases. The linear theory for modeling orographic precipitation~\cite{smith2004linear}, which is an improvement on the upslope model, introduces a time delay component between condensation and rainfall and vertical moisture dynamics. However, the linear theory model is sensitive to its parameters~\cite{paeth2017efficient}. One of the most significant disadvantages of the linear theory model is its inability to account for the spatial variability of the climate variables and parameters. On the other hand, the upslope model incorporates spatial variability of the input variables, which makes it worthwhile despite its biases.
 
\section{Methods}
\label{section:methods}
Our approach downscales low-resolution rainfall data from the European Centre for Medium-Range Weather Forecasts (ECMWF) Reanalysis ~\cite{hersbach2020era5} (ERA5). The high-resolution rainfall fields are comparable to Daymet gridded daily rainfall dataset~\cite{thornton2021gridded}, which serves as the ground truth. We have re-gridded the Daymet rainfall data from a Lambert Conformal Conic projection to an Equirectangular projection with 0.01$^{\circ}$ horizontal resolution, with the help of a radial basis function kernel, to bring it to the same coordinate system as ERA5. Because of model and data collection biases, ERA5 and Daymet precipitation maps do not have one-to-one correspondence at a daily scale, which makes learning a downscaling function between them challenging. A two-step downscaling process overcomes this problem. In the first step (GAN-1), rainfall is downscaled from $0.25^{\circ}$ to $0.1^{\circ}$ resolution, using ERA5 data as the predictor and the corresponding ERA5-Land data as the ground truth. ERA5-land provides a replay of the land surface component of ERA5 at a finer resolution and has high spatial and temporal correspondence with ERA5~\cite{munoz2021era5}. In the second step (GAN-2), upscaled Daymet rainfall fields become predictors to the corresponding high-resolution Daymet fields for training downscaling from $0.1^{\circ}$ to  $0.01^{\circ}$ resolution. In sequence, these two trained models provide the pathway from ERA5 predictors to Daymet-resolution downscaled rainfall. Even though the transformation from ERA5-Land to Daymet may contain bias, our approach corrects it in the final bias correction step. For this study, rainfall events with 90\textsuperscript{th} percentile or above define extreme rainfall. We train and evaluate only using extreme rain events train, discarding all data associated with non-extreme events. Figure~\ref{fig:overview} shows a schematic overview of the proposed methodology, and the following subsections discuss its components.

\subsection{Conditional Gaussian Process}
\label{subsection:gaussianprocess}

We use an iterative conditional Gaussian process (CGP) regressor to build a dynamic data-driven downscaling method. First, we index the training pair of low- and high-resolution rainfall fields on a manifold~\cite{ravela2016dynamically} to query and retrieve nearest neighbors and actively estimate the downscaling function. At each iteration, the downscaled rainfall field estimate upscales again, which targets new data on the manifold for the next learning iteration. Upon convergence, it produces a “first-guess” downscaled rainfall field.

Let a low-resolution rainfall field be $L_{query}$, and its high-resolution counterpart $H_{query}$ to be generated. We make a nearest neighbor search through the low-resolution fields($L_{train}$) of the manifold for the closest match to $L_{query}$, and we denote it as $L_0$. We also obtain the high-resolution counterpart associated with $L_0$, denoted as $H_0$. Now we iteratively improve $L_0$ and $H_0$ until $L_{query}$ and $L_k$ (transformed $L_0$ after $k$\textsuperscript{th} iteration) converges.

\begin{equation}
\label{eqn:gaussianprocessdownscaling}
\begin{aligned}
  & e_k = L_{query} - L_k \\
  & H_{k+1} = H_k + \alpha D(e_k) \\
  & L_{k+1} = U(L_k)
\end{aligned}
\end{equation}

Here $D$ and $U$ are downscaling and upscaling functions, respectively. $\alpha$ is an adjustment rate constant. In this study, we have used bilinear interpolation as the upscaling function and the following Gaussian process regressor as the downscaling function.

\begin{equation}
D(e_k) = {C_{HL} C_{LL}^{-1} e_k} 
\end{equation}

where $C_{LL}$ is the sample conditional covariance of $L_{train}$ neighbors on the manifold, $C_{HL}$ is the cross-covariance between $H_{train}$ and $L_{train}$. To overcome dimensionality issues~\cite{yadav2020machine}, ensemble-based reduced-rank square-root methods~\cite{ravela2007fast, ravela2010realtime} are employed. If the equation \ref{eqn:gaussianprocessdownscaling} converge after $k$\textsuperscript{th} iteration, we assign $H_k$ to our desired “first-guess” downscaled field $H_{query}$.

\subsection{Upslope Orographic Precipitation Estimation}
\label{subsection:orographicprecipitation}

We use the upslope model~\cite{roe2005orographic} for estimating orography-induced precipitation, assuming that precipitation is proportional to the total condensation rate in a vertical column of a saturated atmosphere induced by the vertical wind velocity.
\begin{equation}
\label{eqn:upslopemodel}
   S = - \int_{p_{s}}^{p_{toa}} w \frac{\partial (\rho q_s)}{\partial p} \,dp
\end{equation}
Here $S$ is the orography-induced condensation rate, and $w$ is the orographic vertical wind velocity. Further, $\rho$ is the air density, $q_s$ is the saturation specific humidity, $p$ is atmospheric pressure level, and $p_s$ and $p_{toa}$ are atmospheric pressure at the surface and the top of the atmosphere, respectively. In this study, we assume the top of the atmosphere to be at 200 hPa level. The orography-induced vertical wind velocity at the surface is estimated by,
\begin{equation}
\label{eqn:verticalvelocity}
   w_{s} = \vec{u} \cdot \nabla Z_{e} + \vec{v} \cdot \nabla Z_{n},
\end{equation}
where $\vec{u}$ and $\vec{v}$ are zonal and meridional components of horizontal wind at the surface, and $Z_{e}$ and $Z_{n}$ are the slope of the surface at eastward and northward directions. We interpolate the elevation of the surface to the resolution of ERA5-land before estimating the slope. $w$ is presumed to decrease linearly from the surface to the top of the atmosphere, where it becomes zero. The saturation moisture content (i.e., $\rho q_{s}$) is estimated by,
\begin{equation}
   \rho q_{s} = \frac{R_d}{R_v^2} \frac{e_s(T)}{T}
\end{equation}
where $R_d$ is the gas constant of dry air (287.04 J/Kg/K), $R_v$ is the gas constant of saturated air (461.5 J/Kg/K), $T$ is the air temperature and $e_s$ is the saturation vapor pressure. $e_s$ is estimated by,
\begin{equation}
\label{eqn:saturationvaporpressure}
   e_{s} = e_{s0} \exp \left( \frac{L_v}{R_v} \left( \frac{1}{T_0} - \frac{1}{T} \right) \right),
\end{equation}
where $L_v$ is latent heat of vaporization (2.26 $\times$ 10\textsuperscript{6} J/Kg) and $e_{s0}$ is the reference saturation vapor pressure at the reference temperature $T_0$. When $T_0$ is 273.16 K, $e_{s0}$ is 611 Pa.

Equations \ref{eqn:verticalvelocity}-\ref{eqn:saturationvaporpressure} enable orography-induced vertical wind velocity and saturation moisture content calculations at discrete pressure levels up to 200 hPa, where ERA5 model outcomes are available. We compute the gradient at each pressure level using the second-order finite difference to estimate the integral in the equation \ref{eqn:upslopemodel}.

\subsection{Adversarial Learning}
\label{subsection:gan}

\begin{sidewaysfigure*}
\centering
\includegraphics[width=22cm]{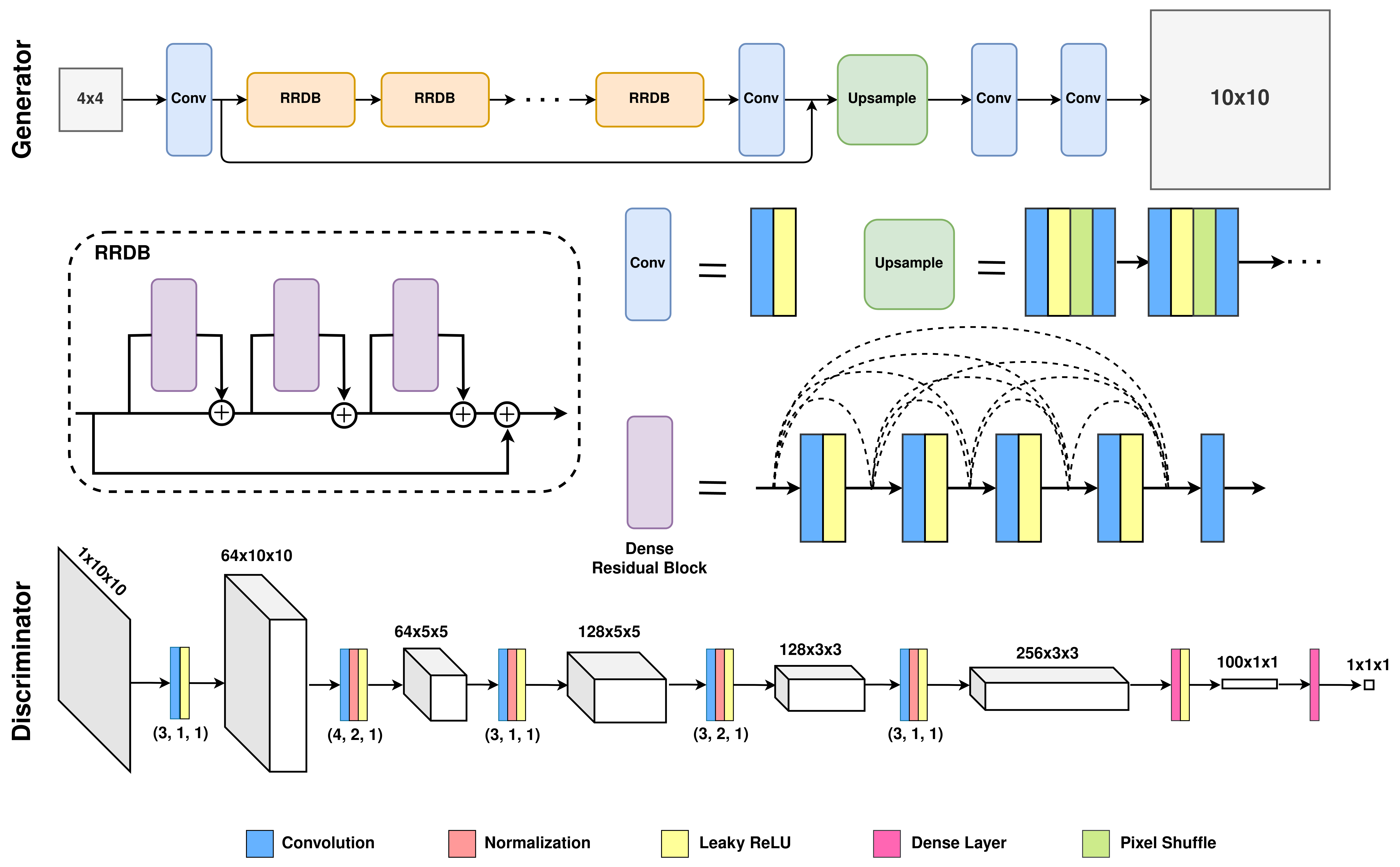}
\caption {Schematic representation of the architecture of the generative adversarial network used in this study. The top depicts the Generator and its components, and the bottom shows the Discriminator.}
\label{fig:ganarchitecture}
\end{sidewaysfigure*}

In Generative Adversarial Networks (GAN), a Generator ($\mathcal{G}$) and a Discriminator ($\mathcal{D}$) network each play a game. Here, the game's outcome is to produce high-resolution rainfall from low-resolution input. The Generator($\mathcal{G}$($L;\alpha_{\mathcal{G}}$)) maps the input low-resolution ($L$) rainfall to a super-resolution reconstruction using a deep convolutional and upsampling network with parameters $\alpha_{\mathcal{G}}$. The Discriminator($\mathcal{D}$($L;\alpha_{\mathcal{D}}$)), repeatedly fed fine-resolution rainfall ground truth ($H$) and the super-resolution generator output, learns to tell them apart. The output of the discriminator is a scalar value, which represents the probability of a rainfall field being “real” (i.e., ground truth). As the two networks train iteratively as adversaries, the Generator's rainfall fields become more realistic, and the Discriminator gets better at distinguishing between them. This is a minimax optimization problem, where the Generator trying to increase $\log \mathcal{D}(\mathcal{G}(L)$, while the Discriminator is trying to reduce it by optimizing their parameters $\alpha_{\mathcal{G}}$ and $\alpha_\mathcal{D}$. The adversarial loss functions of the networks are expressed as the following.~\cite{goodfellow2014generative}
\begin{equation}
\begin{aligned}
    \mathcal{L}_\mathcal{D} (L, H) & =  - \mathbb{E}_H \left[ \log \mathcal{D}(H) \right] - \mathbb{E}_L \left[ 1 - \log \mathcal{D}(\mathcal{G}(L)) \right] \\
    \mathcal{L}_\mathcal{G} (L) & = \mathbb{E}_L \left[ 1 - \log \mathcal{D}(\mathcal{G}(L)) \right]
\end{aligned}
\end{equation}

However, in this study, we are using the Relativistic Average GAN (RaGAN) approach, which compares the discriminator outcome of a real image ($H$) with the average of the fake images ($\mathcal{G}(L)$) and vice versa. Compared to its non-relativistic counterpart, the Relativistic Discriminator increases stability and generates higher-quality samples~\cite{jolicoeur2018relativistic}. The discriminator loss function ($\mathcal{L}_\mathcal{D}$) in our study is given by,
\begin{alignat*}{2}
    \mathcal{L}_\mathcal{D} (L, H) &= &&\mathcal{L}_\mathcal{D}^{real} + \mathcal{L}_\mathcal{D}^{fake}\\
    &= &&- \mathbb{E}_H \left[ \log \left( \mathcal{D}(H) - \mathbb{E}_L \left( \mathcal{D}(\mathcal{G}(L)) \right) \right) \right]\\
    & &&+ \mathbb{E}_L \left[ \log \left( \mathcal{D}(\mathcal{G}(L)) - \mathbb{E}_H \left( \mathcal{D}(H) \right) \right) \right]
    \addtocounter{equation}{1}\tag{\theequation}
\end{alignat*}
The loss function for the Generator ($\mathcal{L}_\mathcal{G}$) considers adversarial Loss as well as pixel-wise $\ell_1$ Loss.
\begin{alignat*}{2}
    \mathcal{L}_\mathcal{G} (L, H) &= &&\mathcal{L}_\mathcal{G}^{adv.} + \lambda \mathcal{L}_\mathcal{G}^{\ell_1} \\
    &= &&- \mathbb{E}_L \left[ \log \left( \mathcal{D}(\mathcal{G}(L)) - \mathbb{E}_H \left( \mathcal{D}(H) \right) \right) \right]\\
    & &&+ \lambda \ell_1({G}(L))
    \addtocounter{equation}{1}\tag{\theequation}
\end{alignat*}
where $\lambda \in \mathbb{R}$ is a regularization factor tunable as a hyperparameter.

The network architectures of the Generator and Discriminator follow  Enhanced Super-Resolution GAN~\cite{wang2018esrgan};  Figure \ref{fig:ganarchitecture} shows a schematic representation. The basic building block of the ResNet-style~\cite{he2016deep} generator is a dense residual block, which is multiple convolutional blocks connected with dense connections. The ESRGAN~\cite{wang2018esrgan} approach replaced them with Residual in Residual Dense Block (RRDB), which consists of stacked dense residual blocks connected with skip connections. The convolutional component of the Generator stacks RRDBs with a global skip connection. It operates on the low-resolution space, followed by an upsampling part that increases the resolution of the rainfall field. In our model, for GAN-1, we have delegated the upsampling job to CGP, and the GAN performs only the convolutional operations on the high-resolution space. GAN-2 does not use CGP, but instead uses a sub-pixel convolution~\cite{shi2016real} (also known as pixel-shuffle)-based upsampling network. For the Discriminator, we have used a VGG-style~\cite{simonyan2014very} deep convolutional network that converts a given real/fake rainfall field into a single value. That is interpretable as the probability that the rainfall field is “real”. Unlike ESRGAN, we do not use any pre-trained VGG network to compute perceptual Loss, as they are unsuitable for capturing climate data features.

\subsection{Risk Quantification}
\label{subsection:riskquantification}
Due to the lack of one-to-one correspondence between ERA5 and Daymet, the final downscaled rainfall and ground truth are not event-wise comparable. However, one can compare the rainfall risk estimated from both of them. A two-parameter Generalized Pareto distribution~\cite{hosking1987parameter} is fit to the annual return periods (R) calculated from the empirical cumulative distribution function (E) of rainfall $r$ from each grid point of each spatial grid and time window of interest.
\begin{equation}
R(r) = \frac{1}{1-E(r)} \\
\end{equation}
The probability density function of the Pareto distribution is given by,
\begin{equation}
      f (r| k, \sigma) = \frac{1}{\sigma} \left( 1 + \frac{kr}{\sigma} \right) ^ {-1-\frac{1}{k}}
\end{equation}
We compare the average annual return period curves simulated by the fitted Pareto distributions for both high-resolution truth and super-resolution predictions and estimate the difference between their mean (bias) and standard error.

\subsection{Stochastic Perturbation and Bias Correction}
\label{subsection:stochasticmodelling}

ERA5 generally underestimates rainfall risk due to its tendency to underestimate the intensity of severe storms, leading to bias. To reduce this bias, we model the observed distribution of rainfall excess beyond the 99.9\textsuperscript{th} percentile, conditioned on spatial location and season (month of the rainfall event). Injecting perturbations from the excess rainfall distribution into the deterministically downscaled rainfall fields improves bias. Additionally, it yields higher-order moments useful for an optimal estimation-based bias correction method (equation \ref{eqn:optimal_estimation}) for additional improvement. We estimate the excess rainfall distribution on the training dataset and develop the bias correction equations on the validation dataset. The optimal estimator is: 
\begin{equation}
\label{eqn:optimal_estimation}
    y_{p}^* = y_{p} + \mathrm{Cov}(y_{p}) \left( \mathrm{Cov}(y_{p}) + \mathrm{Cov}(y_{t}) \right)^{-1}(y_{t} - y_{p})
\end{equation}
Here $y_{t}$, $y_{p}$ and $y_{p}^*$ represents annual return period curves for ground truth, the mean of the stochastic injections (the prior) and bias-corrected (the posterior) rainfall respectively. Again, reduced-rank square-root methods are helpful for highly resolved (and, therefore, high-dimensional) risk curve ensembles~\cite{ravela2007fast, ravela2010realtime}.  We back-project the bias-corrected return period curve using the quantile mapping method to bias-corrected rainfall fields.

\begin{figure*}[htb!]
\centering
\includegraphics[trim=5cm 7cm 0 1cm,scale=0.125] {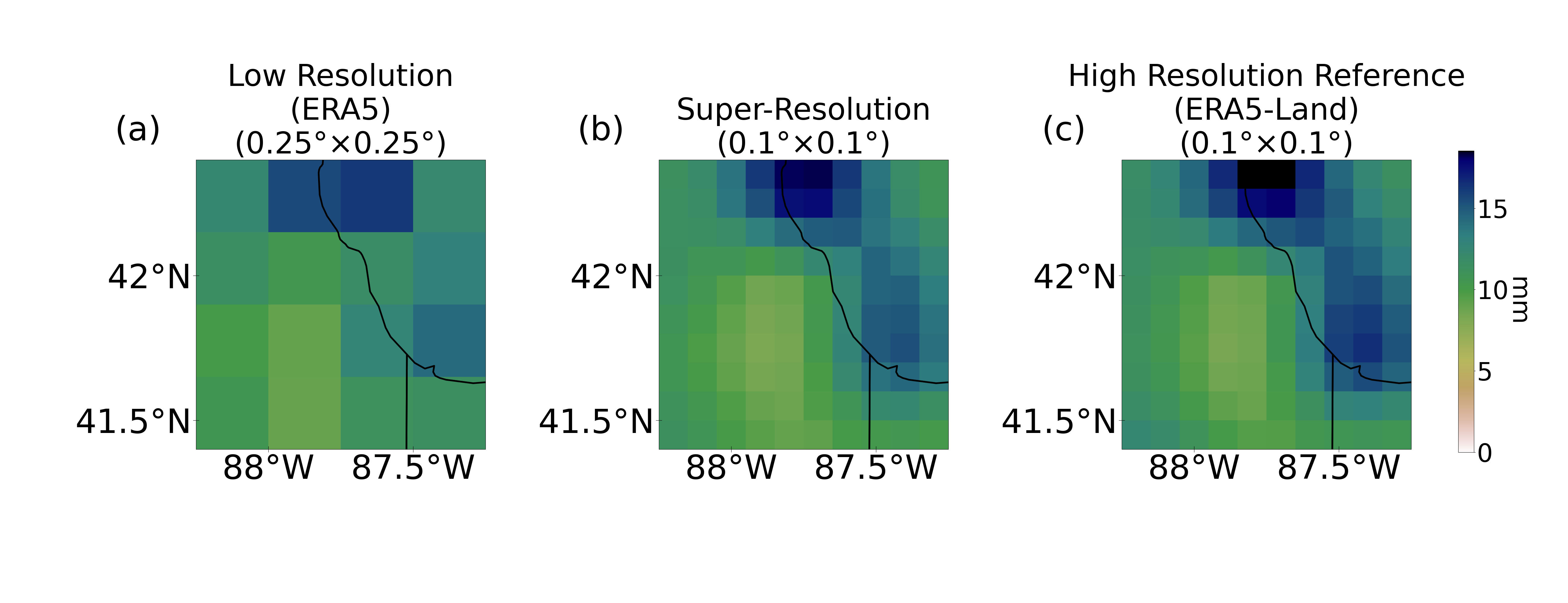}
\caption{Qualitative evaluation of GAN-1 that transforms rainfall from $0.25^{\circ}$ to $0.1^{\circ}$ resolution. Comparison of rainfall fields from \textbf{(a)} ERA5 (low resolution, $0.25^{\circ}$), \textbf{(b)} super-resolution reconstruction ($0.1^{\circ}$), and \textbf{(c)} ERA5-Land (ground truth, $0.1^{\circ}$).}
\label{fig:downscalingera}
\end{figure*}

\begin{figure*}[htb!]
\centering
\includegraphics[trim=5cm 7cm 0 1cm,scale=0.125]{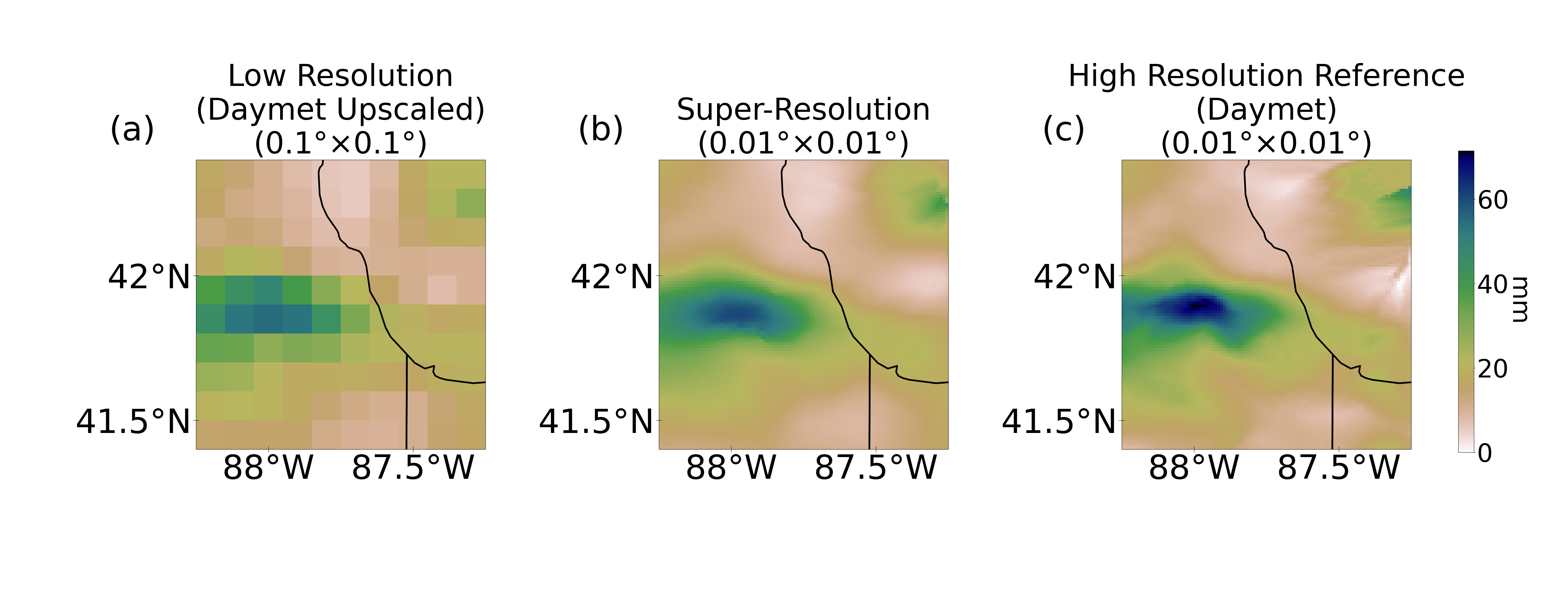}
\caption{Qualitative evaluation of GAN-2 that transforms rainfall from $0.1^{\circ}$ to $0.01^{\circ}$ resolution. Comparison of rainfall fields from \textbf{(a)} Upscaled Daymet (low resolution, $0.1^{\circ}$), \textbf{(b)} super-resolution reconstruction ($0.01^{\circ}$), and \textbf{(c)} Daymet (ground truth, $0.01^{\circ}$).}
\label{fig:downscalingdaymet}
\end{figure*}

\begin{figure*}[htb!]
\centering
\includegraphics[trim=5cm 7cm 0 1cm,scale=0.125]{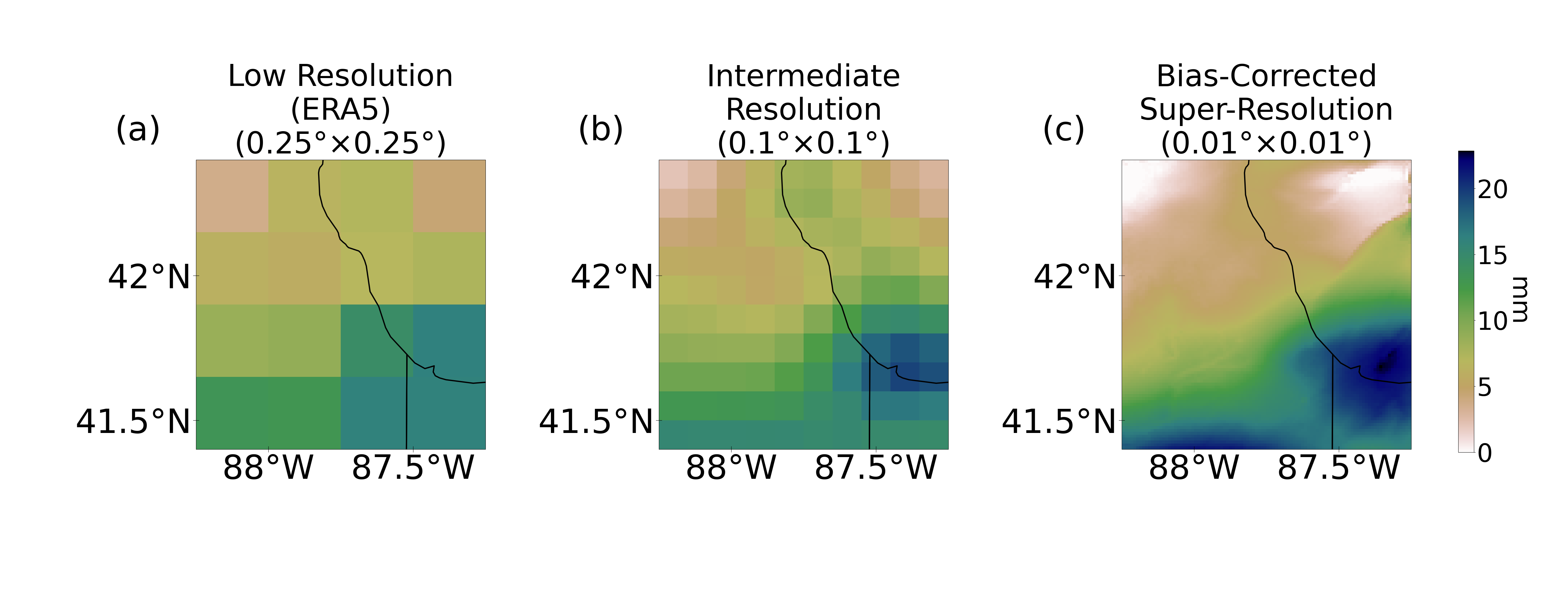}
\caption{Qualitative evaluation of the combined downscaling model that transforms rainfall from $0.25^{\circ}$ to $0.01^{\circ}$ resolution. Comparison of rainfall fields from \textbf{(a)} ERA5 (low resolution, $0.25^{\circ}$), \textbf{(b)} Intermediate resolution reconstruction ($0.1^{\circ}$), and \textbf{(c)} Bias-corrected super-resolution reconstruction ($0.01^{\circ}$).}
\label{fig:downscalingcombined}
\end{figure*}

\section{Results}
\label{section:results}
\begin{figure*}[htbp]
\centering
\includegraphics[trim=3cm 0 0 3cm, scale=0.3]{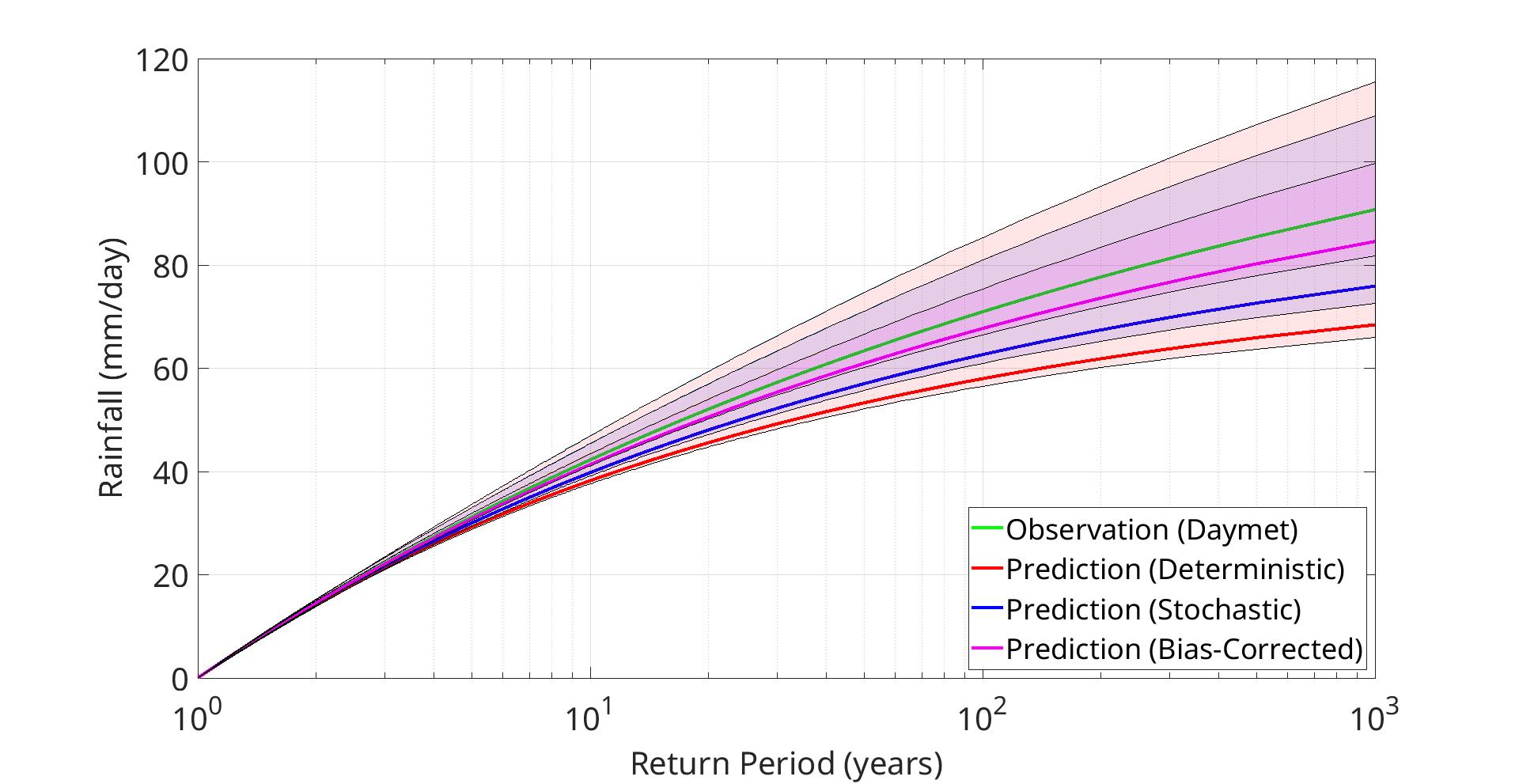}
\caption{Comparison of observed rainfall risk (green) with deterministic (red), stochastic (blue), and bias-corrected (magenta) prediction. The Generalized Pareto distributions fitted to empirically calculated return periods estimate the mean annual rainfall-return period curves. The solid center lines denote mean, and the shaded areas denote standard error.}
\label{fig:downscaling_evaluation}
\end{figure*}

To evaluate the method, the geographic region surrounding Cook County (Chicago), Illinois, USA, a flood-prone area, is selected as the domain of this study. Meteorological data within a $1^{\circ} \times 1^{\circ}$ bounding box between $41.4^{\circ}$N-$42.4^{\circ}$N and $88.25^{\circ}$W-$87.25^{\circ}$W are extracted and divided into three sets: training (1981-1999), validation (2000-2009), and testing (2010-2019). Figure~\ref{fig:downscalingera}-\ref{fig:downscalingdaymet} present qualitative assessments of the performance of the proposed methods (GAN-1 and GAN-2) by comparing low-resolution input, super-resolution output, and high-resolution observation for a particular, extreme event. The downscaled rainfall fields capture the spatial pattern of observed rainfall well. For the combined downscaling model (figure~\ref{fig:downscalingcombined}), a comparison with observation is impossible due to a lack of correspondence between ERA5 and Daymet rainfall. However, we evaluate the model's performance to capture the extreme rainfall risk in the Figure~\ref{fig:downscaling_evaluation}, where we compare the Pareto-distribution simulated mean annual extreme rainfall return period curve for observations, deterministic downscaling, stochastic downscaling, and the optimal estimate. Deterministic predictions underestimate risk, but optimal bias correction substantially closes the gap. The final annual projections show around 6.8\% bias and below 10\% standard error even at an extreme 1000-year return period. It is important to note that the data may not contain a 1000-year rainfall event but is a prediction from the fitted distribution.

\section{Discussions}
\label{section:discussions}
This study leverages data, physics, and machine learning to develop a dynamic approach for rainfall downscaling and estimating extreme rainfall risk. This method applies to the current climate for short-term projections, where the persistence of climate applies. In principle, it is also adaptable to long-term climate change scenarios with a small ensemble of high- and low-resolution climate model runs. In our downscaling method, priming Generative Adversarial Learning with a dynamic data-driven conditional Gaussian Process alleviates the need for extensive data augmentation strategies, given sparse historical extreme rainfall data. Priming with orography also accrues similar advantages; additionally, this enables the physical consistency of the result. We note that Cook County is orographically less challenging than other areas. Experiments are underway to demonstrate the approach for sharper orographic contrasts. Generative adversarial downscaling can capture fine-scale information; however, adapting the Loss function for finer-scale details requires additional work. We also plan to incorporate learning loops to learn the kernels in the dynamic data-driven component and parameter estimation for the physics-based orographic precipitation model, further casting it in a dynamic data-driven framework.

\section*{Acknowledgment}
The authors thank Joaquin Salas and Michael Barbehenn for their review and comments on the manuscript. The authors acknowledge support from Liberty Mutual (029024-00020), ONR (N00014-19-1-2273), and the two MIT Climate Grand Challenge projects on Weather and Climate Extremes and Climate Resilience Early Warning Systems.

\section*{Data Availability}
All datasets used in this study to train the downscaling model are publicly available. Daymet daily rainfall dataset~\cite{thornton2021gridded} is available from the Distributed Active Archive Center for Biogeochemical Dynamics (DAAC) of the Oak Ridge National Laboratory (ORNL), USA. Hourly rainfall, surface and atmospheric temperature, surface meridional and zonal wind speed, and specific humidity from both ERA5~\cite{hersbach2020era5} and ERA5-land~\cite{munoz2019era5} models are downloaded from Copernicus Climate Change Service (C3S) Climate Data Store (CDS) and temporally upscaled to daily timescale. The topographic elevation data of the study area is obtained at 30 m horizontal resolution from the NASADEM Global Digital Elevation Model, which is created by processing interferometric Synthetic Aperture Radar (SAR) data from the Shuttle Radar Topography Mission (SRTM) \cite{crippen2016nasadem}.

\bibliographystyle{IEEEtran}
\bibliography{references}

\begin{thebibliography}{10}
\providecommand{\url}[1]{#1}
\csname url@samestyle\endcsname
\providecommand{\newblock}{\relax}
\providecommand{\bibinfo}[2]{#2}
\providecommand{\BIBentrySTDinterwordspacing}{\spaceskip=0pt\relax}
\providecommand{\BIBentryALTinterwordstretchfactor}{4}
\providecommand{\BIBentryALTinterwordspacing}{\spaceskip=\fontdimen2\font plus
\BIBentryALTinterwordstretchfactor\fontdimen3\font minus
  \fontdimen4\font\relax}
\providecommand{\BIBforeignlanguage}[2]{{%
\expandafter\ifx\csname l@#1\endcsname\relax
\typeout{** WARNING: IEEEtran.bst: No hyphenation pattern has been}%
\typeout{** loaded for the language `#1'. Using the pattern for}%
\typeout{** the default language instead.}%
\else
\language=\csname l@#1\endcsname
\fi
#2}}
\providecommand{\BIBdecl}{\relax}
\BIBdecl

\bibitem{IPCC2021}
S.~Seneviratne, X.~Zhang, M.~Adnan, W.~Badi, C.~Dereczynski, A.~Di~Luca,
  S.~Ghosh, I.~Iskandar, J.~Kossin, S.~Lewis, F.~Otto, I.~Pinto, M.~Satoh,
  S.~Vicente-Serrano, M.~Wehner, and B.~Zhou, \emph{Weather and Climate Extreme
  Events in a Changing Climate}.\hskip 1em plus 0.5em minus 0.4em\relax
  Cambridge, United Kingdom and New York, NY, USA: Cambridge University Press,
  2021, p. 1513–1766.

\bibitem{neumann2015joint}
J.~E. Neumann, K.~Emanuel, S.~Ravela, L.~Ludwig, P.~Kirshen, K.~Bosma, and
  J.~Martinich, ``Joint effects of storm surge and sea-level rise on us coasts:
  new economic estimates of impacts, adaptation, and benefits of mitigation
  policy,'' \emph{Climatic Change}, vol. 129, no.~1, pp. 337--349, 2015.

\bibitem{neumann2015risks}
J.~E. Neumann, K.~A. Emanuel, S.~Ravela, L.~C. Ludwig, and C.~Verly, ``Risks of
  coastal storm surge and the effect of sea level rise in the red river delta,
  vietnam,'' \emph{Sustainability}, vol.~7, no.~6, pp. 6553--6572, 2015.

\bibitem{john2022quantifying}
A.~John, H.~Douville, A.~Ribes, and P.~Yiou, ``Quantifying cmip6 model
  uncertainties in extreme precipitation projections,'' \emph{Weather and
  Climate Extremes}, vol.~36, p. 100435, 2022.

\bibitem{wilbyDownscalingGeneralCirculation1997}
R.~Wilby and T.~Wigley, ``Downscaling general circulation model output: A
  review of methods and limitations,'' \emph{Progress in Physical Geography:
  Earth and Environment}, vol.~21, no.~4, pp. 530--548, Dec. 1997.

\bibitem{trautner2020informative}
M.~Trautner, G.~Margolis, and S.~Ravela, ``Informative neural ensemble kalman
  learning,'' \emph{arXiv:2008.09915}, 2020.

\bibitem{salas22}
J.~Salas, A.~Saha, and S.~Ravela, ``Machine learning approaches to model
  inter-annual flood loss risk from national flood insurance program claims and
  extreme rainfall data,'' \emph{arXiv}, 2022.

\bibitem{dddasbook}
E.~Blasch, S.~Ravela, and A.~Aved, ``Handbook of dynamic data driven
  applications systems,'' \emph{Springer}, p. 750, 2018.

\bibitem{ravela2016dynamically}
S.~Ravela, ``Dynamically deformable resampled random manifolds for
  high-dimensional, nonlinear inference in geoscience in the presence of
  uncertainty,'' in \emph{AGU Fall Meeting Abstracts}, vol. 2016, 2016, pp.
  IN13C--1670.

\bibitem{ravela2007fast}
S.~Ravela and D.~McLaughlin, ``Fast ensemble smoothing,'' \emph{Ocean
  Dynamics}, vol.~57, no.~2, pp. 123--134, 2007.

\bibitem{ravela2010realtime}
S.~Ravela, J.~Marshall, C.~Hill, A.~Wong, and S.~Stransky, ``A realtime
  observatory for laboratory simulation of planetary flows,'' \emph{Experiments
  in fluids}, vol.~48, no.~5, pp. 915--925, 2010.

\bibitem{giorgi2009addressing}
F.~Giorgi, C.~Jones, G.~R. Asrar \emph{et~al.}, ``Addressing climate
  information needs at the regional level: the cordex framework,'' \emph{World
  Meteorological Organization (WMO) Bulletin}, vol.~58, no.~3, p. 175, 2009.

\bibitem{wood2002long}
A.~W. Wood, E.~P. Maurer, A.~Kumar, and D.~P. Lettenmaier, ``Long-range
  experimental hydrologic forecasting for the eastern united states,''
  \emph{Journal of Geophysical Research: Atmospheres}, vol. 107, no. D20, pp.
  ACL--6, 2002.

\bibitem{najafiStatisticalDownscalingPrecipitation2011}
M.~R. Najafi, H.~Moradkhani, and S.~A. Wherry, ``Statistical {{Downscaling}} of
  {{Precipitation Using Machine Learning}} with {{Optimal Predictor
  Selection}},'' \emph{Journal of Hydrologic Engineering}, vol.~16, no.~8, pp.
  650--664, Aug. 2011.

\bibitem{kannanNonparametricKernelRegression2013}
S.~Kannan and S.~Ghosh, ``A nonparametric kernel regression model for
  downscaling multisite daily precipitation in the {{Mahanadi}} basin,''
  \emph{Water Resources Research}, vol.~49, no.~3, pp. 1360--1385, 2013.

\bibitem{mehrotra2005nonparametric}
R.~Mehrotra and A.~Sharma, ``A nonparametric nonhomogeneous hidden markov model
  for downscaling of multisite daily rainfall occurrences,'' \emph{Journal of
  Geophysical Research: Atmospheres}, vol. 110, no. D16, 2005.

\bibitem{zhang2015new}
X.~Zhang and X.~Yan, ``A new statistical precipitation downscaling method with
  bayesian model averaging: a case study in china,'' \emph{Climate Dynamics},
  vol.~45, no.~9, pp. 2541--2555, 2015.

\bibitem{xuDownscalingProjectionMultiCMIP52020}
R.~Xu, N.~Chen, Y.~Chen, and Z.~Chen, ``Downscaling and {{Projection}} of
  {{Multi}}-{{CMIP5 Precipitation Using Machine Learning Methods}} in the
  {{Upper Han River Basin}},'' p. e8680436, Mar. 2020.

\bibitem{schoof2001downscaling}
J.~T. Schoof and S.~C. Pryor, ``Downscaling temperature and precipitation: A
  comparison of regression-based methods and artificial neural networks,''
  \emph{International Journal of Climatology}, vol.~21, no.~7, pp. 773--790,
  2001.

\bibitem{cannonQuantileRegressionNeural2011}
A.~J. Cannon, ``Quantile regression neural networks: {{Implementation}} in
  {{R}} and application to precipitation downscaling,'' \emph{Computers \&
  Geosciences}, vol.~37, no.~9, pp. 1277--1284, Sep. 2011.

\bibitem{heSpatialDownscalingPrecipitation2016}
X.~He, N.~W. Chaney, M.~Schleiss, and J.~Sheffield, ``Spatial downscaling of
  precipitation using adaptable random forests,'' \emph{Water Resources
  Research}, vol.~52, no.~10, pp. 8217--8237, 2016.

\bibitem{tripathiDownscalingPrecipitationClimate2006}
S.~Tripathi, V.~V. Srinivas, and R.~S. Nanjundiah, ``Downscaling of
  precipitation for climate change scenarios: {{A}} support vector machine
  approach,'' \emph{Journal of Hydrology}, vol. 330, no.~3, pp. 621--640, Nov.
  2006.

\bibitem{xu2020precipatch}
M.~Xu, Q.~Liu, D.~Sha, M.~Yu, D.~Q. Duffy, W.~M. Putman, M.~Carroll, T.~Lee,
  and C.~Yang, ``Precipatch: A dictionary-based precipitation downscaling
  method,'' \emph{Remote Sensing}, vol.~12, no.~6, p. 1030, 2020.

\bibitem{sachindraMachineLearningDownscaling2019}
D.~A. Sachindra and S.~Kanae, ``Machine learning for downscaling: The use of
  parallel multiple populations in genetic programming,'' \emph{Stochastic
  Environmental Research and Risk Assessment}, vol.~33, no.~8, pp. 1497--1533,
  Sep. 2019.

\bibitem{wangSequencebasedStatisticalDownscaling2020}
Q.~Wang, J.~Huang, R.~Liu, C.~Men, L.~Guo, Y.~Miao, L.~Jiao, Y.~Wang,
  M.~Shoaib, and X.~Xia, ``Sequence-based statistical downscaling and its
  application to hydrologic simulations based on machine learning and big
  data,'' \emph{Journal of Hydrology}, vol. 586, p. 124875, Jul. 2020.

\bibitem{miaoImprovingMonsoonPrecipitation2019}
Q.~Miao, B.~Pan, H.~Wang, K.~Hsu, and S.~Sorooshian, ``Improving {{Monsoon
  Precipitation Prediction Using Combined Convolutional}} and {{Long Short Term
  Memory Neural Network}},'' \emph{Water}, vol.~11, no.~5, p. 977, May 2019.

\bibitem{vandalIntercomparisonMachineLearning2019}
T.~Vandal, E.~Kodra, and A.~R. Ganguly, ``Intercomparison of machine learning
  methods for statistical downscaling: The case of daily and extreme
  precipitation,'' \emph{Theoretical and Applied Climatology}, vol. 137, no.~1,
  pp. 557--570, Jul. 2019.

\bibitem{shaDeepLearningBasedGriddedDownscaling2020}
Y.~Sha, D.~J.~G. Ii, G.~West, and R.~Stull, ``Deep-{{Learning}}-{{Based Gridded
  Downscaling}} of {{Surface Meteorological Variables}} in {{Complex Terrain}}.
  {{Part II}}: {{Daily Precipitation}},'' \emph{Journal of Applied Meteorology
  and Climatology}, vol.~59, no.~12, pp. 2075--2092, Dec. 2020.

\bibitem{dong2015image}
C.~Dong, C.~C. Loy, K.~He, and X.~Tang, ``Image super-resolution using deep
  convolutional networks,'' \emph{IEEE transactions on pattern analysis and
  machine intelligence}, vol.~38, no.~2, pp. 295--307, 2015.

\bibitem{kim2016accurate}
J.~Kim, J.~K. Lee, and K.~M. Lee, ``Accurate image super-resolution using very
  deep convolutional networks,'' in \emph{Proceedings of the IEEE conference on
  computer vision and pattern recognition}, 2016, pp. 1646--1654.

\bibitem{lim2017enhanced}
B.~Lim, S.~Son, H.~Kim, S.~Nah, and K.~Mu~Lee, ``Enhanced deep residual
  networks for single image super-resolution,'' in \emph{Proceedings of the
  IEEE conference on computer vision and pattern recognition workshops}, 2017,
  pp. 136--144.

\bibitem{haris2018deep}
M.~Haris, G.~Shakhnarovich, and N.~Ukita, ``Deep back-projection networks for
  super-resolution,'' in \emph{Proceedings of the IEEE conference on computer
  vision and pattern recognition}, 2018, pp. 1664--1673.

\bibitem{ledig2017photo}
C.~Ledig, L.~Theis, F.~Husz{\'a}r, J.~Caballero, A.~Cunningham, A.~Acosta,
  A.~Aitken, A.~Tejani, J.~Totz, Z.~Wang \emph{et~al.}, ``Photo-realistic
  single image super-resolution using a generative adversarial network,'' in
  \emph{Proceedings of the IEEE conference on computer vision and pattern
  recognition}, 2017, pp. 4681--4690.

\bibitem{wang2018esrgan}
X.~Wang, K.~Yu, S.~Wu, J.~Gu, Y.~Liu, C.~Dong, Y.~Qiao, and C.~Change~Loy,
  ``Esrgan: Enhanced super-resolution generative adversarial networks,'' in
  \emph{Proceedings of the European conference on computer vision (ECCV)
  workshops}, 2018.

\bibitem{vandalDeepSDGeneratingHigh2017}
\BIBentryALTinterwordspacing
T.~Vandal, E.~Kodra, S.~Ganguly, A.~Michaelis, R.~Nemani, and A.~R. Ganguly,
  ``{{DeepSD}}: {{Generating High Resolution Climate Change Projections}}
  through {{Single Image Super}}-{{Resolution}},'' \emph{arXiv:1703.03126
  [cs]}, Mar. 2017. [Online]. Available: \url{http://arxiv.org/abs/1703.03126}
\BIBentrySTDinterwordspacing

\bibitem{singh2019downscaling}
A.~Singh, A.~Albert, and B.~White, ``Downscaling numerical weather models with
  conditional generative adversarial networks.'' \emph{CLI info}, 2019.

\bibitem{watson2020investigating}
C.~D. Watson, C.~Wang, T.~Lynar, and K.~Weldemariam, ``Investigating two
  super-resolution methods for downscaling precipitation: Esrgan and car,''
  \emph{arXiv preprint arXiv:2012.01233}, 2020.

\bibitem{roe2005orographic}
G.~Roe, ``Orographic precipitation,'' \emph{Annual Review of Earth and
  Planetary Sciences}, vol.~33, pp. 645--671, 2005.

\bibitem{collier1975representation}
C.~Collier, ``A representation of the effects of topography on surface rainfall
  within moving baroclinic disturbances,'' \emph{Quarterly Journal of the Royal
  Meteorological Society}, vol. 101, no. 429, pp. 407--422, 1975.

\bibitem{sinclair1994diagnostic}
M.~R. Sinclair, ``A diagnostic model for estimating orographic precipitation,''
  \emph{Journal of Applied Meteorology and Climatology}, vol.~33, no.~10, pp.
  1163--1175, 1994.

\bibitem{smith2004linear}
R.~B. Smith and I.~Barstad, ``A linear theory of orographic precipitation,''
  \emph{Journal of the Atmospheric Sciences}, vol.~61, no.~12, pp. 1377--1391,
  2004.

\bibitem{paeth2017efficient}
H.~Paeth, F.~Pollinger, H.~M{\"a}chel, C.~Figura, S.~Wahl, C.~Ohlwein, and
  A.~Hense, ``An efficient model approach for very high resolution orographic
  precipitation,'' \emph{Quarterly Journal of the Royal Meteorological
  Society}, vol. 143, no. 706, pp. 2221--2234, 2017.

\bibitem{hersbach2020era5}
H.~Hersbach, B.~Bell, P.~Berrisford, S.~Hirahara, A.~Hor{\'a}nyi,
  J.~Mu{\~n}oz-Sabater, J.~Nicolas, C.~Peubey, R.~Radu, and D.~Schepers, ``{The
  ERA5 global reanalysis},'' \emph{Quarterly Journal of the Royal
  Meteorological Society}, vol. 146, no. 730, pp. 1999--2049, 2020.

\bibitem{thornton2021gridded}
P.~E. Thornton, R.~Shrestha, M.~Thornton, S.-C. Kao, Y.~Wei, and B.~E. Wilson,
  ``{Gridded daily weather data for North America with comprehensive
  uncertainty quantification},'' \emph{Sci. Data}, vol.~8, no.~1, pp. 1--17,
  2021.

\bibitem{munoz2021era5}
J.~Mu{\~n}oz-Sabater, E.~Dutra, A.~Agust{\'\i}-Panareda, C.~Albergel,
  G.~Arduini, G.~Balsamo, S.~Boussetta, M.~Choulga, S.~Harrigan, H.~Hersbach,
  B.~Martens, D.~G. Miralles, M.~Piles, N.~J. Rodr{\'\i}guez-Fern{\'a}ndez,
  E.~Zsoter, C.~Buontempo, and J.-N. Th{\'e}paut, ``Era5-land: a
  state-of-the-art global reanalysis dataset for land applications,''
  \emph{Earth System Science Data}, vol.~13, no.~9, p. 4349–4383, 2021.

\bibitem{yadav2020machine}
N.~Yadav, S.~Ravela, and A.~R. Ganguly, ``Machine learning for robust
  identification of complex nonlinear dynamical systems: Applications to earth
  systems modeling,'' \emph{arXiv:2008.05590}, 2020.

\bibitem{goodfellow2014generative}
I.~Goodfellow, J.~Pouget-Abadie, M.~Mirza, B.~Xu, D.~Warde-Farley, S.~Ozair,
  A.~Courville, and Y.~Bengio, ``Generative adversarial nets,'' in
  \emph{Advances in Neural Information Processing Systems}, vol.~27, 2014.

\bibitem{jolicoeur2018relativistic}
A.~Jolicoeur-Martineau, ``{The relativistic discriminator: a key element
  missing from standard GAN},'' \emph{arXiv:1807.00734}, 2018.

\bibitem{he2016deep}
K.~He, X.~Zhang, S.~Ren, and J.~Sun, ``Deep residual learning for image
  recognition,'' in \emph{Proceedings of the IEEE conference on computer vision
  and pattern recognition}, 2016, pp. 770--778.

\bibitem{shi2016real}
W.~Shi, J.~Caballero, F.~Husz{\'a}r, J.~Totz, A.~P. Aitken, R.~Bishop,
  D.~Rueckert, and Z.~Wang, ``Real-time single image and video super-resolution
  using an efficient sub-pixel convolutional neural network,'' in
  \emph{Proceedings of the IEEE conference on computer vision and pattern
  recognition}, 2016, pp. 1874--1883.

\bibitem{simonyan2014very}
K.~Simonyan and A.~Zisserman, ``Very deep convolutional networks for
  large-scale image recognition,'' \emph{arXiv:1409.1556}, 2014.

\bibitem{hosking1987parameter}
J.~Hosking and J.~Wallis, ``Parameter and quantile estimation for the
  generalized pareto distribution,'' \emph{Technometrics}, vol.~29, no.~3, pp.
  339--349, 1987.

\bibitem{munoz2019era5}
J.~Mu{\~n}oz-Sabater, ``Era5-land hourly data from 1981 to present,''
  \emph{Copernicus Climate Change Service (C3S) Climate Data Store (CDS)},
  2019.

\bibitem{crippen2016nasadem}
R.~Crippen, S.~Buckley, E.~Belz, E.~Gurrola, S.~Hensley, M.~Kobrick,
  M.~Lavalle, J.~Martin, M.~Neumann, Q.~Nguyen, P.~Rosen, J.~Shimada,
  M.~Simard, and W.~Tung, ``Nasadem global elevation model: Methods and
  progress,'' \emph{International Archives of the Photogrammetry, Remote
  Sensing and Spatial Information Sciences — ISPRS Archives}, vol. XLI-B4,
  2016.

\end{thebibliography}
\end{document}